\documentclass[usegraphicx,usenatbib,useapjfonts,apj]{emulateapj}
\shorttitle{Mass-metallicity gradient}
\shortauthors{Spolaor et al.}

\begin{document}

\title{The mass-metallicity gradient relation of early-type galaxies\altaffilmark{1}}

\author{Max Spolaor\altaffilmark{2}, Robert N. Proctor, Duncan A. Forbes, Warrick J. Couch}
\affil{Centre for Astrophysics \& Supercomputing, Swinburne University of Technology, Hawthorn, VIC 3122, Australia}

\altaffiltext{1}{Based on observations obtained at the Gemini observatory, which is operated by the Association of Universities for Research in Astronomy, Inc., under a cooperative agreement with the NSF in behalf of the Gemini partnership: the National Science Foundation (United States), the Particle Physics and Astronomy Research Council (United Kingdom), the National Research Council (Canada), CONICYT (Chile), the Australian Research Council (Australia), CNPq (Brazil), and CONICET (Argentina).}
\altaffiltext{2}{Corresponding Author: mspolaor@astro.swin.edu.au}
\begin{abstract}
We present a newly observed relation between galaxy mass and radial metallicity gradients of early-type galaxies. Our sample of 51 early-type galaxies encompasses a comprehensive mass range from dwarf to brightest cluster galaxies. The metallicity gradients are measured out to one effective radius by comparing nearly all of the Lick absorption-line indices to recent models of single stellar populations. 
The relation shows very different behavior at low and high masses, with a sharp transition being seen at a mass of $\sim 3.5\times10^{10}$~M$_{\odot}$ (velocity dispersion of $\sim$~140~km~s$^{-1}$, M$_{B} \sim-19$). Low-mass galaxies form a tight relation with mass, such that metallicity gradients become shallower with decreasing mass and positive at the very low-mass end. Above the mass transition point several massive galaxies have steeper gradients, but a clear downturn is visible marked by a broad scatter. The results are interpreted in comparison with competing model predictions. We find that an early star-forming collapse could have acted as the main mechanism for the formation of low-mass galaxies, with star formation efficiency increasing with galactic mass. The high-mass downturn could be a consequence of merging and the observed larger scatter a natural result of different merger properties. These results suggest that galaxies above the mass threshold of $\sim 3.5\times10^{10}$~M$_{\odot}$ might have formed initially by mergers of gas-rich disk galaxies and then subsequently evolved via dry merger events. The varying efficiency of the dissipative merger-induced starburst and feedback processes have shaped the radial metallicity gradients in these high-mass systems.
\end{abstract}


\section{Introduction}
Radial metallicity gradients have been found to characterize the stellar population of early-type galaxies. They are the chemodynamical fossil imprints of galaxy formation and evolution mechanisms. The presence of such population gradients and their relationship to galaxy structural parameters provides a strong constraint on galaxy formation. Competing scenarios that describe the formation of early-type galaxies predict contrasting behaviors for the mass-metallicity gradient relation. 

In the classical models of monolithic collapse, stars form during a rapid dissipative collapse while the gas sinks to the center of the forming galaxy (\citealt{larson74}; \citealt{larson75}; \citealt{carlberg84}; \citealt{arimoto87}). The inflowing gas is chemically enriched by evolving stars, and contributes metal-rich fuel for star formation. Thus a negative radial gradient is established, in which central stars are more metal-rich than those born in outer galaxy regions. Metallicity-dependent gas cooling and a time delay in the occurrence of galactic winds (i.e.~onsets that vary with the local escape velocity) cooperate in steepening any metallicity gradients (\citealt{matteucci94}; \citealt{martinelli98}; \citealt{pipino08}). The predicted metallicity gradients are as steep as -0.35 (\citealt{larson74}), -1.0 (\citealt{larson75}), and -0.5 dex per radius dex (\citealt{carlberg84}). The depth of the galaxy potential well regulates the effectiveness of these processes, so that a strong correlation between metallicity gradient and galaxy mass is expected. High-mass galaxies retain more metals allowing steep negative metallicity gradients of -0.5 dex per radius dex, compared to low-mass systems with almost zero gradient (\citealt{gibson97}; \citealt{chiosi02}; \citealt{kawata03}).

In the hierarchical clustering scenario of galaxy formation, early-type galaxies are the product of merging events (\citealt{kauff93}; \citealt{kauff98}; \citealt{lucia06}; \citealt{lucia07}). The turbulent mixing caused by merging events is expected to ``wash out'' any possible ambient gradient presents in the progenitor galaxies (\citealt{white80}). Simulations suggest that stellar population and kinematic properties of a merger remnant are determined by the progenitors' mass ratio and by the amount of gas involved in the merger event (e.g., \citealt{naab06}). Metallicity gradients are produced by a rapid central merger-induced starburst and their strength is proportional to the efficiency of the dissipative process (\citealt{hopk08}). The gradients are predicted to only weakly depend on the remnant galactic mass (\citealt{bekki99}; \citealt{koba04}; \citealt{hopk08b}).

Metallicity gradients in early-type galaxies have been studied by means of increasing strength of metal-sensitive lines and the reddening of colors. Unfortunately, most estimates are made under the crucial a priori assumption that this increase is entirely caused by a radial variation of stellar metallicity and therefore neglecting age effects. Typically, past works have derived total metallicities using just a few metal-lines (i.e.~$H\beta$, $Mg_{2}$, $Fe5270$) and have sampled the relatively central galactic region inside 0.5 effective radii. To date, a clear and consistent picture of the behavior of the mass-metallicity gradient relation has not emerged from observational studies.

In a sample of 42 early-type galaxies, \cite{carollo93} observed that their $Mg_{2}$ gradients became shallower for galaxies with masses less than about 10$^{11}$~M$_{\odot}$, while more massive systems show no correlation. From a literature compilation, \cite{koba99} studied metal-line index gradients of 80 elliptical galaxies without finding any statistically significant trend with mass. Recently, \cite{ogando05} and \cite{forbes05} claimed opposite trends for a sample of 35 and 29 massive early-type galaxies. \cite{ogando05} showed that both flattening and steepening of metallicity gradients are possible in massive galaxies. On the other hand, \cite{forbes05} found that gradients become steeper in more massive systems.

Here we present new results from high signal-to-noise Gemini GMOS observations of 14 low-mass early-type galaxies in the Virgo and Fornax clusters (M.~Spolaor et al., in preparation). By using all of the Lick indices and adopting the technique of \cite{proctor02}, we have been able to break the age-metallicity degeneracy and obtain reliable metallicity gradient measurements out to more than two effective radii. Combining our measurements with those from a novel literature compilation, we extend our study to a broad mass range (i.e., from dwarfs to brightest cluster galaxies). The final sample of metallicity gradients in 51 early-type galaxies is a uniquely uniform sample due to the use of nearly all of the Lick indices and the steps taken to separate the age-metallicity degeneracy. 

In this Letter, we are able to probe for the first time the low-mass end of the mass-metallicity gradient relation. A clear picture of the relation along the full mass range is finally possible. Its interpretation in the context of competing model predictions allows us to place constraints on galaxy formation mechanisms.

\section{A consistent data sample}
Long-slit spectroscopic observations of 14 dwarf Virgo and Fornax cluster early-type galaxies were performed with GMOS at the Gemini South telescope during the semesters 2006B (Program ID GS-2006B-Q-74) and 2008A (Program ID GS-2008A-Q-3). The galaxies were selected to be low-mass using the central stellar velocity dispersion ($\sigma$) and the total B-band absolute magnitude (M$_{B}$) as independent proxies of galaxy mass. The sample uniformly covers the range $1.6 <\log\sigma < 2.15$ and $-19.5 <$~M$_{B} < -16$. The spectrograph slit was set to 0.5~arcsec and covered 5.5~arcmin in length. We used the 600~line~mm$^{-1}$ grism blazed at 5250~\AA~providing a dispersion of 0.458~\AA~per~pixel and a spatial resolution of 0.1454~arcsec~per~pixel. The spectroscopic data were reduced using the Gemini data processing software. We spatially resolved several apertures out to one effective radius along the major axis of each galaxy. The spatial width (i.e.~the number of CCD rows binned) for each extracted aperture increases with radius to achieve a signal-to-noise ratio of $\sim 30$ at the central wavelength of 5200~\AA. The spectra cover the full range of the 21 line-strength indices of the original Lick/IDS (Image Dissector Scanner) system (\citealt{worthey94}) plus the additional age-sensitive Balmer indices $H\delta_{A}$, $H\gamma_{A}$, $H\delta_{F}$ and $H\gamma_{F}$ (\citealt{worthey97}). The measured indices were calibrated to the Lick system as described in Appendix A of \cite{spola08b}. We note that a comprehensive analysis of line-strength indices of these new data are the subject of a future work by M.~Spolaor et al.~(in preparation).

\begin{figure*}
\centering
\includegraphics[scale=0.85]{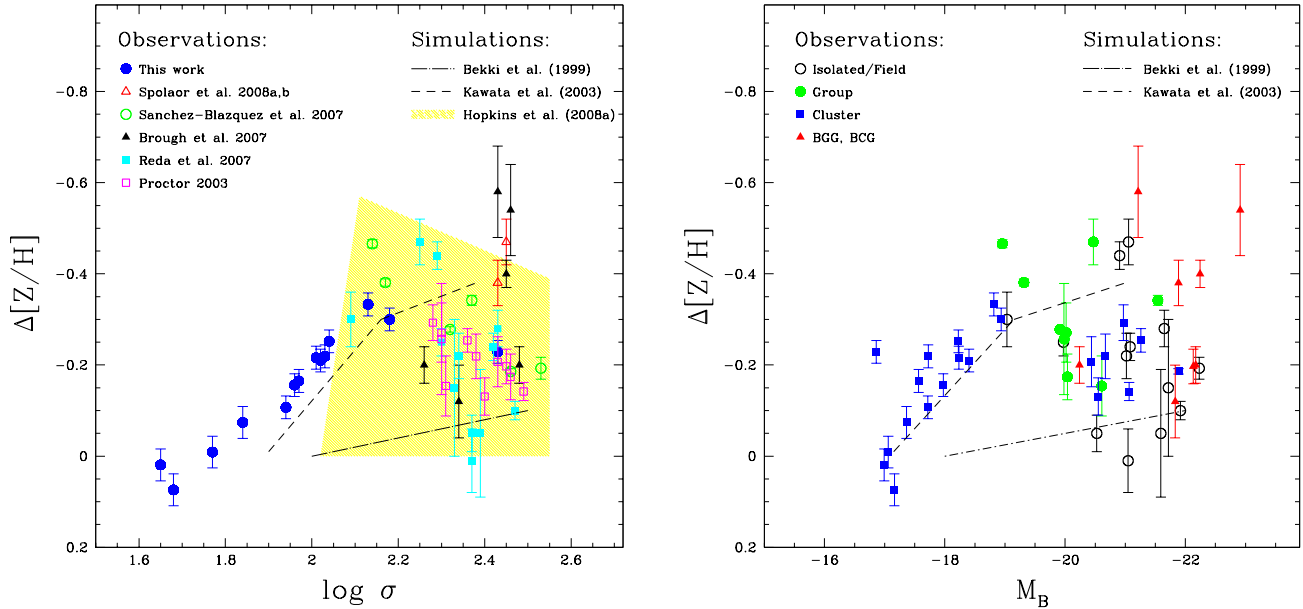}
\caption[]{Metallicity gradients $\Delta[Z/H]$ as a function of galaxy central velocity dispersion $\log\sigma$ (left panel) and total B-band absolute magnitude M$_{B}$ (right panel), both independent proxies of galaxy mass. In the left panel, the 51 data points are coded by original data sample. In the right panel, the points are coded by environment: isolated/field, group, cluster and brightest group/cluster galaxies (BGG, BCG). Also shown are the mass dependent predictions from the merging models of \cite{bekki99} as dot-dashed lines, and the dissipative collapse models of \cite{kawata03} as dashed lines. The region occupied by the remnants of major mergers between gas-rich disk galaxies, as simulated by \cite{hopk08}, is shown by the yellow shading.}
\label{merge}
\end{figure*}

Independent radial profiles of luminosity-weighted age, total metallicity [Z/H] and $\alpha$-abundance ratio [$\alpha$/Fe] were derived from the fully calibrated indices. Briefly, the technique of \cite{proctor02} is a statistical $\chi^{2}$ minimization of the deviation between the observed and modeled index values as a fraction of the index errors. The strength of the method is that it works with as many indices as possible in order to break the age-metallicity degeneracy that affects each index differently. Consistent with the other studies, we used the single stellar population models of \cite{thomas04}, which include the variation of Lick indices to abundance ratios. 

Gradients were obtained by a linear least-squares fit to the radial metallicity profile weighted by the errors associated with each metallicity value. Therefore, metallicity gradients express the derivative variation of the total metallicity along the galactocentric radius (scaled by the effective radius). We represent the values in logarithmic space as $\Delta [Z/H] / \Delta\log (R/R_{e})$ (hereafter $\Delta[Z/H]$). To be able to compare gradients from different data samples we uniformly fit the data points beyond the seeing limit (i.e., $\geq$ 1 arcsec) out to one effective radius. We note that the inclusion of these very inner regions introduces only a statistically insignificant change in the gradients.

Our full sample is a compilation of 51 metallicity gradients in early-type galaxies from six different studies that adopted the same criteria and analysis technique. Spolaor et al.~(2008a, 2008b) studied the star formation and chemical evolutionary history of the two brightest galaxies of the NGC~1407 group. \cite{pat07} used Keck LRIS long-slit spectra to study a sample of 11 early-type galaxies covering a wide range in mass, situated in the field, poor groups and the Virgo cluster. \cite{brough07} focused on the stellar population and kinematic properties of three nearby brightest group galaxies and 3 brightest cluster galaxies located at z~$\sim$~0.055. \cite{reda07} analyzed the metallicity gradients of 12 isolated luminous early-type galaxies. The sample of Proctor (2003) consists of 11 early-type galaxies located in the Leo cloud and the Virgo cluster. We note that five of the 11 galaxies of the \cite{pat07} sample overlap with the Proctor (2003) sample; their gradients are similar, however we adopted the Proctor values due to his use of all 25 Lick indices versus the 20 used by \cite{pat07}. 

\section{Results and Discussion}
We measured the central stellar velocity dispersion for our galaxy sample by averaging the dispersion values within a radius of $R_{e}/8$. Total B-band absolute magnitudes were estimated from total apparent magnitudes given in the NED database and corrected for Galactic extinction. For each galaxy we assumed the distance modulus from surface brightness fluctuations of \cite{tonry01}  and \cite{mei05}. We applied the correction of \cite{jensen03}  to the values of Tonry.

In Figure~1 we present the relation between metallicity gradients and two independent proxies of galaxy mass. Gradient values of the 51 galaxies are shown as a function of $\sigma$ (left panel), coded by the data source, and M$_{B}$ (right panel), coded by environment. We defined four environmental classes: isolated/field, group, cluster and brightest group/cluster galaxies (BGGs, BCGs). The results are compared to theoretical predictions of the competing galaxy formation models of \cite{bekki99} (dot-dashed black line) and \cite{kawata03} (dashed black line). For completeness, we also show the simulations of \cite{hopk08} of gas-rich major merger remnants which have  ``excess'' central light due to a merger-induced central starburst. The region they occupy is shown by the yellow shading.

The models of \cite{bekki99} simulated the formation of early-type galaxies by merging gas-rich disk galaxies of varying mass ratio at redshift z~$>$~2. The chemodynamical evolution of mass-dependent galaxy properties was investigated by inclusion of supernova feedback and time delay between star formation and metal ejection. As can be seen in Figure~1, the metallicity gradients found in merger remnants are shallow and only weakly dependent on the remnant galaxy mass. 

Nevertheless, \cite{hopk08} have been able to reproduce merger remnants with a variety of metallicity gradients (0.0~$\leq \Delta [Z/H] \leq$~$-$0.6) in simulations of gas-rich disk galaxies major mergers, including feedback processes from galactic winds and black hole growth (\citealt{springel05}). Similarly, they did not find a strong dependence with mass but they stressed that the degree of dissipation in the central merger-induced starburst is the most important factor in determining the gradient strength in the remnant. The impact on stellar population gradients of possible subsequent dry mergers of these systems has the same effect observed for passively evolved gas-rich merger remnants, in which metallicity gradients tend to flatten with time (\citealt{hopk08b}).

In the chemodynamical simulations of \cite{kawata03}, an elliptical galaxy begins via the dissipative collapse of a slow-rotating, overdense, top-hat protogalactic gas cloud, on which CDM perturbations are superimposed. They find a mass-dependence of both the supernova-driven galactic winds and the gas infall rate, underlying their important role in creating and shaping metallicity gradients. As a result, metallicity gradients are steeper in higher mass galaxies than those of lower mass systems. 

In Figure~1 our new results populate the previously unexplored low-mass branch of the parameter space, while the literature values sample the high-mass region. Galaxies with $\log\sigma < 2.15$ (M$_{B} \geq -19$) exhibit a tight trend in which the metallicity gradient becomes less negative and hence shallower with decreasing mass. At the very low-mass end ($\log\sigma < 1.8$, M$_{B} \geq -17.5$) gradient values change sign becoming positive or null. The predictions of \cite{kawata03} appear to describe the observed trend, suggesting an early star-forming collapse as the main mechanism for the formation of low-mass galaxies. Specifically, star formation efficiency increases with galactic mass.

A significantly different behavior can be seen in galaxies with $\log\sigma > 2.15$ (M$_{B} \leq -19$). A downturn from the tight low-mass trend is visible marked by a broad scatter such that galaxies with increasing mass are characterized by shallower metallicity gradients. However, some of the most massive galaxies classified as BGGs and BCGs have steeper negative gradients. The downturn and the broad scatter, could be a consequence of merging and of the amount of gas involved in the events, which characterize the formation of more massive galaxies. If a significant fraction of gas dissipates in a rapid central starburst, then, the metallicity gradient of the remnant will reflect the mean difference between that of the old pre-merger population and that set by the newly formed young stars. The gradients of the pre-merger populations have been partially flattened by violent relaxation, which acts most strongly at larger galactocentric radii. The metallicity gradient in the remnant is expected, initially, to be steep and then to weaken slightly over $\sim$~3~Gyr after the merger (\citealt{hopk08}).

\begin{figure}
\centering
\includegraphics[scale=0.42]{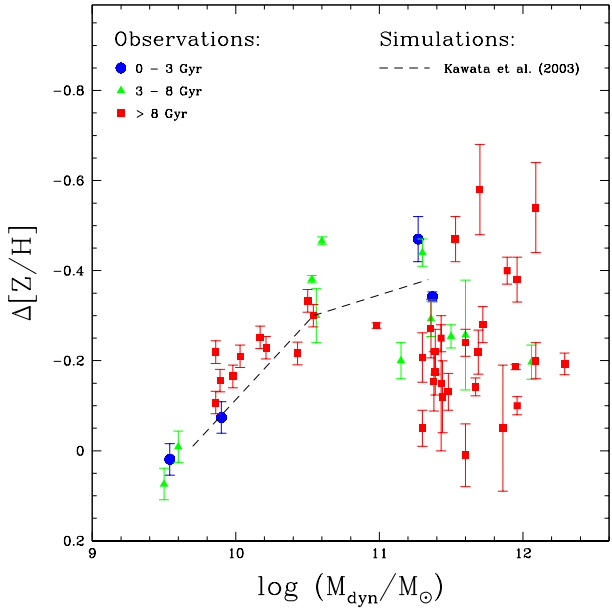}
\caption{Metallicity gradients $\Delta[Z/H]$ as a function of galaxy dynamical mass M$_{dyn}$.  The points are coded by the central age value of the galaxies: young ($0 -3$ Gyr), intermediate ($3-8$ Gyr) and old age ($ > 8$ Gyr). Also shown are the mass-metallicity gradient predictions from the dissipative collapse models of \cite{kawata03} as dashed lines. The models of \cite{bekki99} and \cite{hopk08} are not available.}
\label{agemass}
\end{figure}

In Figure~2 we show metallicity gradient values as a function of dynamical mass. We estimated the dynamical mass of the galaxies by $\log($M$_{dyn}) = 2\log(\sigma) + \log(R_{e}) + 3.1$, where $R_{e}$ is the effective radius expressed in pc. By averaging the ages estimated from the models within a radius of $R_{e}/8$ we obtained the galaxy central age. In Figure~2  we coded the points by assigning three age bins: $0-3$ Gyr, $3-8$ Gyr, $> 8$ Gyr. 

The tight low-mass end of the relation is populated by an almost even number of galaxies with young/intermediate ($0-8$ Gyr) and old central ages ($> 8$ Gyr). The galaxies with younger ages have positive or null gradients. Above the high-mass turnover the number of galaxies with old central age becomes prominent. We quantified the mass transition point around the dynamical mass of $\sim 3.5\times10^{10}$M$_{\odot}$, which is similar to the value of $10^{11}$M$_{\odot}$ found in the data of \cite{carollo93}. We note that the mass value of $\sim 3\times10^{10}$M$_{\odot}$ is recurrent in literature as critical transition mass for physical properties of early-type galaxies, such as feedback processes from active galactic nuclei (e.g., \citealt{kauff03}; \citealt{croton06}; \citealt{catt08}).

A broader interpretation is that processes of both formation scenarios arise in the CDM framework for galaxies more massive than $\sim 3.5\times10^{10}$~M$_{\odot}$ ($\log\sigma > 2.15$, M$_{B} \leq -19$). A rapid star-forming collapse might have acted at primordial times creating low-mass progenitor galaxies which have then merged to form today's early-type galaxies. Therefore, merger events are the main mechanism acting to grow galaxy mass and in shaping the subsequent evolution of galaxies above this mass threshold. This has already been suggested by the theoretical work of \cite{koba04}. In her chemodynamical simulations the metallicity gradients do not correlate with the galaxy mass but can be used, in principle, to infer the galaxy merging history.

The significant increase in scatter observed beyond the high-mass turnover could be caused by the different degree of dissipation that characterize merger-induced starbursts and therefore the strength of the remnants' metallicity gradients. A further contribution to the scatter might come from a time-evolution of the merger-induced metallicity gradients.

\section{Conclusions}
For the first time we have probed the relation between galaxy mass and radial metallicity gradients of early-type galaxies. Our results show a sharp transition between the low and high-mass regimes of the relation at a galaxy mass of $\sim 3.5\times10^{10}$~M$_{\odot}$ ($\log\sigma \sim2.15$, M$_{B} \sim-19$).

Galaxies with masses smaller than this transition mass are observed to form a tight relation. Their metallicity gradients become less negative and hence shallower with decreasing mass. The interpretation of the results in comparison with theoretical models suggests that these low-mass galaxies formed during an early star-forming collapse, with star formation efficiency increasing with galactic mass.

Above the high-mass turnover of $\sim 3.5\times10^{10}$~M$_{\odot}$ several massive galaxies have steeper gradients, but a clear downturn from the tight low-mass trend is visible marked by a broad scatter such that gradients become shallower with increasing mass. The downturn could be a consequence of merging, and the observed scatter a natural result of merger properties. The results suggest that galaxies formed by gas-rich mergers and then evolved, growing their mass via subsequent dry merger events. The merger-induced starburst and feedback processes are responsible for creating the radial metallicity gradients which are conserved during dry mergers, but tend to flatten with time.

\acknowledgments
This research has made use of the NASA/IPAC Extragalactic Database (NED). We thank D.~Croton, A.~Graham, G.~K.~T. Hau, T.~Mendel and S.~Burke-Spolaor for their careful reading of this manuscript and many useful discussions. D.F. and R.P. acknowledge the ARC for financial support. We also thank the anonymous referee for his/her constructive comments.

\end{document}